\documentclass[preprint2]{aastex}

\input epsf.sty

\newcommand{\etal}{et~al.\ }
\newcommand{\eg}{e.g.\ }
\newcommand{\ie}{i.e.\ }
\newcommand{\Msun}{M_{\odot}}

\newcommand{\kms}{km~s$^{-1}$}
\newcommand{\ergs}{erg~s$^{-1}$}

\newcommand{\OI}{O~{\sc i}}

\newcommand{\SII}{S~{\sc ii}}

\newcommand{\SiII}{Si~{\sc ii}}

\newcommand{\CaII}{Ca~{\sc ii}}
\newcommand{\TiII}{Ti~{\sc ii}}

\newcommand{\FeII}{Fe~{\sc ii}}

\newcommand{\CoII}{Co~{\sc ii}}

\newcommand{\Nifs}{$^{56}$Ni}
\newcommand{\Mej}{$M_{\rm ej}$}
\newcommand{\KE}{$E$}

\begin{document}

\title{Models for the Type Ic Hypernova SN 2003lw associated with GRB 031203
  \begin{center}
  \bigskip
  {\rm\em Accepted for publication in ApJ}
  \end{center}
}

\author{
Paolo A.~Mazzali\altaffilmark{1,2,3,4},
Jinsong~Deng\altaffilmark{5,1,2},
Elena~Pian\altaffilmark{4},
Daniele~Malesani\altaffilmark{6},
Nozomu~Tominaga\altaffilmark{1,2},
Keiichi~Maeda\altaffilmark{7},
Ken'ichi~Nomoto\altaffilmark{1,2},
Guido Chincarini\altaffilmark{8,9},
Stefano Covino\altaffilmark{8},
Massimo~Della Valle\altaffilmark{10,11},
Dino Fugazza\altaffilmark{8},
Gianpiero Tagliaferri\altaffilmark{8},
Avishay~Gal-Yam\altaffilmark{12,13}
}

\altaffiltext{1}{Department of Astronomy, University of Tokyo,
  Bunkyo-ku, Tokyo 113-0033, Japan}
\altaffiltext{2}{Research Center for the Early Universe, University of Tokyo,
  Bunkyo-ku, Tokyo 113-0033, Japan}
\altaffiltext{3}{Max-Planck-Institut f\"ur Astrophysik,
  Karl-Schwarzschildstr.\ 1, 85748 Garching, Germany}
\altaffiltext{4}{National Institute for Astrophysics--OATs, Via Tiepolo, 11,
  34131 Trieste, Italy}
\altaffiltext{5}{National Astronomical Observatories, CAS, 20A Datun Road,
  Chaoyang District, Beijing 100012, China}
\altaffiltext{6}{International School for Advanced Studies (SISSA),
 via Beirut 2-4, I-34014 Trieste, Italy}
\altaffiltext{7}{Department of Earth Science and Astronomy,
 Graduate School of Arts and Science, University of Tokyo,
 Meguro-ku, Tokyo 153-8902, Japan}
\altaffiltext{8}{National Institute for Astrophysics--OABrera,
 via Bianchi 46, I-23807 Merate (Lc), Italy}
\altaffiltext{9}{Universit\`a di Milano-Bicocca,
Dipartimento di Fisica, piazza delle Scienze 3, I-20126 Milano, Italy}
\altaffiltext{10}{National Institute for Astrophysics--OAArcetri,
 largo Fermi 5, I-50125 Firenze, Italy}
\altaffiltext{11}{European Southern Observatory,
 Karl-Schwarzschildstr.\ 2, 85748 Garching, Germany}
\altaffiltext{12}{Dept. of Astronomy, California Institute of Technology,
 Pasadena, CA 91125}
\altaffiltext{13}{Hubble Fellow}

\begin{abstract}
The Gamma-Ray Burst 031203 at a redshift $z = 0.1055$ revealed a highly
reddened Type Ic Supernova, SN~2003lw, in its afterglow light. This is the
third well established case of a link between a long-duration GRB and a Type Ic
SN. The SN light curve is obtained subtracting the galaxy contribution and is
modelled together with two spectra at near-maximum epochs. A red VLT grism 150I
spectrum of the SN near peak is used to extend the spectral coverage, and in
particular to constrain the uncertain reddening, the most likely value for
which is $E_{G+H}(B-V) \simeq 1.07\pm0.05$. Accounting for reddening,
SN~2003lw is $\sim 0.3$\,mag brighter than the prototypical GRB-SN 1998bw.
Light curve models yield a \Nifs\ mass of $\sim 0.55\Msun$.  The optimal
explosion model is somewhat more massive (\Mej~$\sim 13 \Msun$) and
energetic (\KE~$\sim 6 \times 10^{52}$\,erg) than the model for SN~1998bw,
implying a massive progenitor ($40 - 50 \Msun$).  The mass at high velocity is
not very large ($1.4 \Msun$ above 30000\,\kms, but only $0.1\Msun$ above
60000\,\kms), but is sufficient to cause the observed broad lines. The
similarity of SNe\,2003lw and 1998bw and the weakness of their related GRBs,
GRB031203 and GRB980425, suggest that both GRBs may be normal events viewed
slightly off-axis or a weaker but possibly more frequent type of GRB.
\end{abstract}


\section{Introduction}

The first indication of a connection between long-duration Gamma-Ray Bursts
(GRB) and Type Ic Supernovae (SNe~Ic) was the discovery of a bright SN~Ic,
SN~1998bw, in spatial and temporal coincidence with the nearby GRB~980425
\citep[$z = 0.0085$,][]{gal98}. The spectra of SN~1998bw showed broad P-Cygni
lines of elements such as Fe, Ca, and Si, indicating expansion velocities of
$\sim 0.1 c$.  Models of the spectra and the light curve yielded an isotropic
equivalent expansion kinetic energy \KE $\sim 5 \times 10^{52}$\,erg
\citep{iwa98} and an ejected mass \Mej~$\sim 11 \Msun$, suggesting that
SN~1998bw was the highly energetic explosion of a massive stellar core. The
explosion synthesised a large mass of \Nifs\ for a core-collapse event ($\sim
0.5 \Msun$). The progenitor of SN~1998bw was probably a very massive star
(M$_{\rm ZAMS} \sim 40 \Msun$), and the remnant very likely a black hole. SNe
with broad spectral features, indicative of a large \KE, have been called
``hypernovae''.

Nebular-phase spectra of SN~1998bw were dominated by a strong  [\OI] 6300,
6363\AA\ emission, as is typical of SNe~Ic, but they also showed strong lines
of [\FeII], reflecting the high \Nifs\ production. Another peculiarity was that
the [\OI] line was narrower than the [\FeII] lines. This is not expected in the
spherically symmetric explosion of a CO core \citep{maz01}, and was interpreted
as the result of the ejection of most \Nifs\ along a preferential (polar)
direction in an aspherical explosion \citep{mae02}, reminiscent of the
collapsar model for GRBs \citep{McFW99}. Doubts were raised about the
association of SN~1998bw and GRB~980425 because of the unusual weakness of the
GRB and the apparent presence of an X-ray transient inconsistent in position
with the SN \citep{pian00}, but the reality of this transient was disproved by
observations with both XMM \citep{pian04} and Chandra \citep{kou04}.

The second clear case of association between a GRB and a SN was that of
GRB~030329 and SN~2003dh. This was also a rather nearby GRB ($z = 0.1687$), but
in this case it was a normal one. After extraction from the strong afterglow
light, the SN turned out to be very similar to SN~1998bw
\citep{sta03,hjo03,mat03,lip04}. A study of the light curve and the spectra
confirmed that SN~2003dh was also a very energetic SN~Ic, with overall
properties similar to those of SN~1998bw, although perhaps somewhat less
extreme \citep[\KE $\sim 3.5 \times 10^{52}$\,erg, \Mej $\sim 8 \Msun$,
M(\Nifs)$ \sim 0.35 \Msun$, M$_{ZAMS} \sim 35 \Msun$;][]{maz03,deng05}. A
late-time spectrum of SN~2003dh also shows a narrow [\OI] and broad [\FeII]
lines (Bersier \etal, in prep.).


Given this evidence, it could be expected that the second closest GRB ever
detected (GRB~031203 at $z = 0.1055$) should also reveal SN signatures. The GRB
was distinctly weaker than average, but stronger than 980425. Very strong
reddening and a bright galaxy background made the detection of the SN very
difficult. Nevertheless, a SN was eventually detected photometrically
\citep{tag04,tho04,gal04,cobb04} and spectroscopically \citep[][hereafter
MTC04]{mal04}. Perhaps not surprisingly, SN~2003lw showed a spectrum similar to
those of SNe~1998bw and 2003dh.

The determination of the properties of SN~2003lw requires reliable measurements
of its light curve and spectra. This is made difficult by contamination from
host galaxy light. Therefore, we have reobserved the field after the SN faded
in order to quantify the contribution of the host galaxy and subtract it out
properly.

In this paper we present newly reduced data for SN~2003lw, derive the SN light
curve and model it and the spectra. Using highly energetic SN~Ic models similar
to those used for SN~1998bw we determine the most likely reddening and derive
the properties of SN~2003lw.

\section{Late-time observations}

In order to perform the subtraction of the host galaxy light as accurately as
possible, late-time photometry\footnote{Based on observations performed at
ESO-Paranal under programme ID 073.D-0255.} and spectroscopy of the host galaxy
were performed on 2004 May 22 ($\approx 5$ months after the GRB) with the ESO
VLT-UT1 (Antu) and FORS\,2, at a time when the SN contribution to the total
light should be negligible. The photometry is reported in Tab.~\ref{tb:phot}.
Data reduction and analysis were performed following standard procedures, using
the \texttt{Eclipse} and \texttt{GAIA} packages. The SN aperture photometry was
calibrated using new field photometry of several standard fields. Photometric
analysis shows only a slight fading ($0.04 \pm 0.02$~mag in the $R$ band) with
respect to 2004 March (MTC04).

Spectroscopy was performed with grism 150I ($6000 - 10000$\,\AA). The net
exposure time was 45 minutes, and the seeing was $0.6\arcsec$.  The data were
reduced using standard \texttt{IRAF} routines.  This new "galaxy only" spectrum
was used to remove the host contribution from a red 150I grism spectrum
acquired on 2003 Dec 20, which was not presented in MTC04 because of
uncertainties in host galaxy subtraction. Flux calibration was achieved
observing two standard stars, LTT\,3864 and LTT\,3218. Consistent results were
obtained for the two cases. In order to ensure relative calibration of the
spectra of December 2003 and May 2004 we computed the flux of the galaxy
emission lines in the different observations and introduced a small correction
to make them match exactly. The entire set of VLT spectroscopic observations of
SN~2003lw is reported in Tab.~\ref{tb:spec}.

\section{The bolometric light curve of SN 2003lw}

In order to derive the properties of SN~2003lw through modelling, it is
necessary to construct the bolometric light curve from the observations. This
requires careful subtraction of the host galaxy contribution. Photometry was in
fact dominated by the host light, so that the shape and brightness of the
reduced SN LC are sensitive to the adopted host brightness. Reliable host
photometry can only be obtained when the SN has faded sufficiently. Our new
data were obtained sufficiently late that we can assume this is the case.

We adopted as host galaxy magnitudes the $R$ and $I$ measurements obtained on
2004 May 22 (see Table~1). These fluxes are both lower than those measured in
March in these bands (MTC04), but the difference is at most $2\sigma$,
indicating that the SN contributed by no more than 3\% to the integrated flux
in these bands already in 2004 March \citep[as a comparison, SN~1998bw became 1
mag fainter over the same time interval,][]{pat01}.

As for the host $V$-band magnitude, we had to base our estimate on the last
available $V$-band measurement, which was taken only $\sim 3$ months after the
SN explosion ($V = 20.54 \pm 0.05$, 2004 March 2) and is probably still
affected by the SN. We therefore considered that the SN+host magnitude declined
by 0.2 mag over the 2 months between late Dec 2003 and 2004 Mar 2. A SN~Ic
similar to SN~1998bw declines by $\sim 2.2$\,mag over that time interval. This
suggests that SN~2003lw still makes a contribution of $\sim 3$\% to the total
light on 2004 Mar 2.  Therefore we adopted a value $V({\rm host}) \simeq
20.57\pm0.05$.

Since NIR templates for SN~1998bw are not available, we derived a host
magnitude starting from the magnitudes measured in Feb 2004 (MTC04), and
corrected them as above assuming that the SN makes a contribution in $J$, $H$,
and $K$ similar to that in the $V$ band. Accordingly, we use host galaxy
magnitudes $J=18.28\pm0.05$, $H=17.58\pm0.04$, and $K = 16.69\pm0.06$. These
values are in fair agreement with those reported by \citet{pro04},
\citet{gal04}, and \citet{cobb04}.

One of the major uncertainties in the calibration of the photometry of
SN~2003lw is the reddening. This is large, and may have both a Galactic and a
host component. \citet{sch98} report for the direction of GRB031203/SN~2003lw a
galactic extinction $E_G(B-V) = 1.04$. \citet{pro04} note that this value may
be highly uncertain and suggest a value $E_G(B-V) = 0.78$ as a lower limit.

Spectrum synthesis favours a total reddening $E_{G+H}(B-V) = 1.07 \pm 0.05$
(see \S 5). This is $\sim 0.1$ smaller than the value obtained by
\citet{pro04}, and is formally consistent with the Galactic value of Schlegel
\etal (1998) and no host extinction. Lack of significant extinction in GRB-host
galaxies is indeed often reported \citep[\eg][and references therein]{wat06b}.
However, following \citep{pro04} we adopt $E_G(B-V) = 0.78$ and $E_H(B-V) =
0.25$, which results from the difference between  $E_{G+H}(B-V) = 1.07$ and
$E_G(B-V) = 0.78$.  Given that extinction laws of GRB hosts are uncertain and
that an SMC-type law may be favoured with respect to a Milky Way-type  law
\citep[see][]{wat06b}, we computed the intrinsic extinction suffered by  the SN
light under the 2 assumptions of a MW-type  \citep{CCM89} and an SMC-type
extinction curve \citep{pei92} in the NIR and optical ranges.  For $E_H(B-V) =
0.25$ the difference in the amount of extinction is at most 5\% in these
wavelength ranges, which is well within our uncertainties.  Therefore we
adopted a MW extinction curve. The value of $E_H(B-V)$ we use here is likely to
represent the maximum possible value.

In order to evaluate the bolometric light curve of SN~2003lw we then proceeded
as follows. First, we verified that SNe 2003lw and 1998bw have similar
spectra.  We constructed the dereddened spectral energy distribution (SED) of
SN~2003lw at every epoch corresponding to an $R$ and/or $I$-band measurement
using the available simultaneous or quasi-simultaneous optical and NIR
photometry. We interpolated the dereddened SEDs of SN~1998bw (using $E(B-V) =
0.016$ from \citet{nak01}) to the phases of the SEDs of SN~2003lw, taking into
account the 1.1 time stretch factor. The SEDs of SN~2003lw are shown in Figure
2 together with the reconstructed SEDs of SN~1998bw.  In order to match the two
SNe in the optical, the SEDs of SN~1998bw were scaled up by a factor 1.3. This
represents the difference in intrinsic luminosity between the two SNe.

Since the data of SN~1998bw do not extend to the IR, we extrapolated linearly
the red part of the optical SED. The agreement between the SEDs of the two SNe
is very good, except for the $K$-band point at the first epoch.  Therefore,
since the SEDs of SN~1998bw extend to bluer wavelengths than those of
SN~2003lw, the {\em ``uvoir''} bolometric magnitudes of SN~2003lw were computed
by integrating the flux under the SN1998bw templates between rest frame 3600
and 22200\,\AA.

The integrated {\em ''uvoir''} LCs of the two SNe are shown in Figure 3. The LC
of SN~2003lw is broader than that of SN~1998bw by a factor of $\sim 1.1$, but
it is very similar to it in shape and is more luminous by $\sim 0.3$\,mag. This
is a smaller difference than what was quoted in MTC04 and \citet{tho04}, and it
is mostly the result of adopting a smaller reddening to SN~2003lw.
Considering the significant systematic uncertainties ($\sim 0.15$\,mag from the
reddening to both SNe, $\sim 0.10$\,mag from possible peculiar velocities
affecting the distance to SN~1998bw) the combined systematic error on the
relative luminosity is $\sim 0.2$\,mag.

\section{Modelling the SN Light Curve}

We based synthetic LC calculations on the 1-D SN LC synthesis code
\citep{iwa00} that was used to model other Type Ic SNe (e.g.
\citealt{nak01,maz02,maz03}). The code solves the energy and momentum
equations of the radiation plus gas in the co-moving frame, and is accurate to
first order in $v/c$. Electron densities and the electron scattering opacity
are determined from the Saha-Boltzmann equation.  We adopted the approximation
proposed by \citet{gom96} for the Eddington factors, and fitted the TOPS
opacities \citep{mag95} to find an empirical relationship between the Rosseland
mean and the electron scattering opacity. The energy deposition from
radioactive decays was calculated with a gray $\gamma$-ray transfer code,
assuming an absorptive opacity of $0.05 Y_e$ cm$^{2}$ g$^{-1}$ \citep{swa95}.
The $\gamma$-ray energy source function was evaluated from the latest nuclear
data \citep{fir99}.

To test the above assumptions as well as other modifications we made to the
code, we computed the LC of SN~1998bw using model CO138E50, which fits the
spectra and LC of SN~1998bw near the LC peak \citep{nak01}. Model CO138E50 has
\Mej$= 10.2 \Msun$ and \KE$ = 5 \times 10^{52}$ erg. We also used the
two-component model developed by \citet{mae03}.  This model has a denser core
than CO138E50 to mimic the effect of a 2-D explosion, and thus it can reproduce
better the late-time LC of SN~1998bw. It has \Mej$= 10.7 \Msun$ and \KE$ = 4.6
\times 10^{52}$ erg. We updated the SN~1998bw bolometric LC of \citet{pat01} to
a distance modulus $\mu=32.76$, corresponding to $H_0 =
72$\,km\,s$^{-1}$\,Mpc$^{-1}$.  Both models require a \Nifs\ mass of $0.41 \pm
0.05 \Msun$. The distribution of \Nifs\ is $0.09 \Msun$ below 5,000\,\kms,
$0.30 \Msun$ between 5,000 and 23,000\,\kms, and $0.02 \Msun$ above
23,000\,\kms\ for CO138E50 and 0.07, 0.32, and $0.02 \Msun$ in the same three
zones for the two-component model. These values are consistent with those of
\citet{nak01}, once the difference in the adopted distance is considered. The
early portion of the LC of SN~1998bw is well reproduced by both models (Figure
4). However, as expected, the two-component model fits the late-time LC better
than CO138E50, confirming the results of \citet{mae03}.

As we showed above, the shapes of the bolometric LCs of SNe 1998bw and 2003lw
are very similar, but the latter is broader by a factor 1.1. It is well known
that the time scale of the LC near the peak, $\tau$, depends on the ejected
mass \Mej, the kinetic energy \KE, and the opacity $\kappa$ as $\tau \propto
\kappa^{1/2}$\Mej$^{3/4}$\KE$^{-1/4}$ \citep{arn82}. Since the difference
between the two SNe is small, we based our calculations on model CO138E50,
increasing its density in order to reproduce the broader LC of SN~2003lw.
Given the relation above, we increased \Mej, and consequently \KE, by a factor
1.2, leading to \Mej$= 12.25 \Msun$ and \KE$ = 6 \times 10^{52}$ erg.

We took the mass and the distribution of ejected \Nifs\ as free parameters, and
assumed that the rest of ejecta consists of 90\% O and 10\% Si by mass.  This
simplified composition was adopted because the two spectra we have available
for modelling are too early to constrain the composition of the bulk of the
ejecta, and also because our model is an approximate 1-D reproduction of the
probably aspherical ejecta. Additionally, we tested the two-component model of
SN~1998bw, also scaled up in mass by a factor 1.2, so that the two-component
model for SN~2003lw has \Mej$= 12.85 \Msun$ and \KE$ = 5.5 \times 10^{52}$ erg.

Modelling the light curve alone does not yield a unique result: simultaneous
modelling of the spectra is required, and this is performed in the next
section. An additional complication in the case of SN~2003lw is the uncertain
reddening. We used a total reddening $E_{G+H}(B-V) = 1.07$, derived from
spectral models as discussed in Section 5.

The shape of the bolometric LC of SN~2003lw is not much affected by the choice
of reddening. Our main model parameters, \Mej\ and \KE, affect the LC shape and
do not change for different reddenings. The overall brightness of course does
change, but it can be reproduced using different amounts of ejected $^{56}$Ni.

Figure \ref{fig5} shows the bolometric LC of SN~2003lw and the best-fitting
synthetic LCs obtained scaling CO138E50 and adjusting the \Nifs\ distribution.
The total \Nifs\ mass is $\sim 0.54\Msun$. As in the case of SN~1998bw, we
adopted a distribution of \Nifs\ with an inner zone at $v<5,000$\,\kms\ and an
outer zone between 5,000 and 23,000\,\kms, and assumed uniform \Nifs\ abundance
within each zone. The best-fitting \Nifs\ distribution is $0.12 \Msun$ in the
inner zone and $0.42 \Msun$ in the outer zone. Unlike the case of SN~1998bw,
lack of early photometry phase makes it impossible to tell whether \Nifs\ was
mixed to the highest velocities, so we did not introduce \Nifs\ above
23,000\,\kms.

The synthetic LC of the scaled two-component model is also shown in Figure
\ref{fig5}. The \Nifs\ mass in this model is also $0.54 \Msun$, of which $0.10
\Msun$ below $v<5,000$\,\kms\ and $0.44 \Msun$ above.  The two-component model
fits better than CO138E50 after maximum, confirming that the dense core of
these models is efficient in absorbing $\gamma$-rays and influences the LC
after peak. Since the $\gamma$-ray deposition efficiency is increased by adding
the inner dense core, a \Nifs\ mass similar to that used for the rescaled
CO138E50 model is sufficient to sustain the higher integrated luminosity of the
LCs of the two-component models.

From the above studies, we cannot select between scaled CO138E50 and the scaled
two-component model. The synthetic LCs differ before maximum, but we do not
know the exact shape of the pre-maximum SN~2003lw LC. Also, this phase is very
sensitive to the \Nifs\ distribution.  Moreover, the two spectra available are
too early to show evidence of a dense core, unlike the cases of other
previously studied HNe
\citep[SNe~1997ef, 1998bw, and 2002ap;][]{min00,mae03,maz02}, and the light
curve coverage does not extend to sufficiently late times to apply the
2-component models of \citet{mae03}. It is therefore unfortunately impossible
to apply the 2D models of the spectra and the light curve as was possible for
SN~1998bw \citep{mae02}. We would however expect that since SN~2003lw was
similar to SN~1998bw its late time properties would also be best described by a
highly aspherical explosion viewed close to the jet axis.

\section{Spectral Modelling}

Because of the brightness of the host galaxy it was only possible to obtain SN
spectra at two epochs near maximum: 2003 December 20 and 30, corresponding to
rest-frame epochs of 16 and 24 days past explosion, respectively.  Here we add
the red 150I grism VLT spectrum of 2003 Dec 20 to the two blue 300V grism
spectra shown in MTC04. This is useful because it is less sensitive to
reddening.  It shows a very strong P-Cygni profile, probably a blend of \CaII\
IR and \OI\ 7772\AA, a typical hypernova signature \citep[\eg][]{maz02}.

We computed synthetic spectra with our Montecarlo code \citep{m&l93,l99,maz00}
for these two epochs. Following the results of the light curve modelling, we
used model CO138E50 rescaled in mass by a factor $f = 1.2$. We did not use the
rescaled two-component model, as it only differs in the innermost part, which
is not probed by the spectra near maximum.  Our models can be used to constrain
the total reddening, and have some leverage on the relative Galactic and host
reddening since the two galaxies are separated by $z = 0.1055$. We adopted
$E_G(B-V) = 0.8$ as a lower limit, attributed the remaining reddening to the
host, and tested different values of rest-frame $E_H(B-V)$.

In Figure 6 we show the December 20 spectrum and 4 synthetic spectra, computed
for $E_H(B-V) = 0.2$, 0.25, 0.3, and 0.35. The main input parameters of the
models and the properties of the synthetic spectra are summarised in Table 3.
Both Galactic and host reddening were computed using the \citet{CCM89} law.
Obviously, the model luminosity increases with increasing reddening. In all
models the composition is dominated by oxygen ($\sim 70$\% by mass) and neon
($\sim 20$\%). \Nifs\ is only $\sim 2$--3\% by mass, and silicon is even less,
$\sim 1$--2\%. This composition is slightly different from that used for light
curve modelling. Including both Ne and O is important for the spectrum
synthesis, while the synthetic LC is not significantly affected by replacing Ne
with O. As reddening decreases, the abundance of \Nifs\ also decreases, while
that of oxygen increases, resulting in a progressively stronger \OI\ absorption
near 7000\AA.

All models are very similar in the blue, where the main absorption features are
due to \FeII, \CoII, and \TiII\ lines (the big trough at 4000-4500\AA), and in
the $V$ region, which is dominated by lines of \SiII\ and \SII. However, the
behaviour in the red, where the strong \CaII-\OI\ P-Cygni line shows
prominently, is rather different. The model with $E_H(B-V) = 0.2$ shows  too
much absorption in this feature. This is a consequence of the high abundance of
oxygen. In this model, in fact, the abundances of the heavier elements that
give rise to the lines in the blue are lower than they are in models with
higher reddening, while the abundance of oxygen is higher.  The models with
$E_H(B-V) = 0.25$, 0.3, and 0.35, all reproduce the \CaII--\OI\ absorption
well, but the model with the highest reddening has a large luminosity, and
consequently a large photospheric velocity, so that the spectral lines are too
red, and is of lower quality.

Models for the December 30 spectrum are shown in Figure 7. Because of the
smaller wavelength coverage at this epoch, it is not easy to discriminate among
the models, all of which look reasonably good. The input values are listed in
Table 4.  The abundances are similar to those used for the December 20
spectrum, but \Nifs\ (and decay products) is reduced by a factor of $\sim 2$.
This shows that more \Nifs\ was produced at higher velocities, contributing to
the earlier spectrum, and confirms the scenario of an aspherical explosion. It
is unfortunate that the earliest spectrum is only at peak, otherwise we may
have been able to determine more accurately the \Nifs\ abundance in the
outermost, fastest moving layers. All models fit the blue part of the spectrum
nicely, but those with $E_H(B-V) = 0.25$ and 0.3 seem to be the best in the
\OI-\CaII\ region, at least if we trust the red end of the spectrum. As was the
case for the December 20 models, the oxygen abundance is higher for models with
lower reddening, causing the exceedingly strong \OI\ absorption near 7200\AA.
Therefore, given our assumed Galactic reddening, $E_G(B-V) = 0.8$, we favour a
value of the host reddening $E_H(B-V) \sim 0.25 \pm 0.05$. This results in a
total reddening of $E_{H+G}(B-V) \sim 1.07 \pm 0.05$ when the redshift of the
host is taken into account. Since the total reddening is the quantity that is
best constrained by our models, our result is also consistent with the Schlegel
\etal (1998) galactic reddening and no host extinction.  With this value of the
reddening the absolute magnitude of SN~2003lw on 2003 Dec 20 is -19.03, which
is consistent with our LC analysis (Table 3).

\section{Discussion}

The properties of SN~2003lw are very similar to those of the other
well-observed GRB-SNe. Given our chosen values of the host reddening, the
rescaled explosion model that best fits both the LC and the spectra of
SN~2003lw has an \KE$ \sim 6 \times 10^{52}$\,erg, \Mej~$ \sim 13 \Msun$. From
the value of \Mej, the inferred mass of the progenitor of SN~2003lw is $\sim
40-50 \Msun$. These values are similar to, although somewhat larger than those
of SN~1998bw. SN~2003lw follows the positive correlation between \KE, M(\Nifs),
and progenitor mass that appears to hold for SNe~Ic \citep{nom04,nom05}. As in
the case of SN~1998bw, SN~2003lw is also likely to have left behind a black
hole remnant.

Figure 8 shows the time evolution of the photospheric velocity as determined
from spectral modelling for the three well-studied GRB-SNe and for other
broad-lined SNe~Ic that did not show an accompanying GRB. The three GRB-SNe
have the highest values, SN~2003dh being somewhat higher than the other two.
Both SNe~Ic SNe~1997ef and 2002ap have much lower photospheric velocities.

It must be stressed that the estimated values of the explosion parameters of
SN~2003lw are necessarily approximate. As we discussed, the estimated peak
brightness of the SN, and hence the derived \Nifs\ mass, depends sensitively on
the uncertain galaxy subtraction and the assumed reddening. Moreover, the
estimate of \KE\ is based on spherically symmetric models of the explosion. If
asymmetries were taken into account, most likely in the form af an aspherical
explosion, as was possible for the much better studied case of SN~1998bw, it is
very likely that the value of \KE\ would be reduced. In the case of SN~1998bw
an estimate based on a spherical model suggested that the true kinetic energy
is smaller by about a factor of 5 with respect to \KE, yielding a true \KE\
$\sim 10^{52}$erg \citep{mae02}. Such a study was made possible for SN~1998bw
by the availability of late-time spectrophotometry, which unfortunately cannot
be obtained for SN~2003lw since the bright host galaxy outshines the SN. The
relation between the true and isotropic equivalent energies depends sensitively
on both the asphericity of the explosion and our viewing angle. If these were
not too different for the three GRB-SNe, as is likely, then the true \KE\ of
SN~2003lw may be reduced by a similar factor, to $\sim 1.2 \times
10^{52}$\,erg, and that of SN~2003dh to $\sim 7 \times 10^{51}$erg.  Even after
such a major reduction, the \KE\ of the three GRB-SNe remains very large
compared to that of typical SNe, deserving them the name ``hypernovae''.

Perhaps the most remarkable aspect is that although the properties of the three
GRB-SNe vary by at most $\pm 30$\%, those of the related GRBs ($\gamma$- and
X-ray energy output) cover $\sim 4$ orders of magnitude, GRB030329 being almost
as energetic as a normal GRB, GRB031203 a factor of 100 lower, and GRB980425 a
factor of 100 lower still.  The durations, $\gamma$-ray fluences and relative
energy ranges for the 3 GRBs are reported in Table 5.  The $\gamma$-ray outputs
of the 3 GRBs (Col. 7) were computed according to
\begin{equation}
 E = \frac{4 \pi d^2 S_\gamma}{1 + z}.
\end{equation}
In the above equation $d$ is the distance, $S_\gamma$ the fluence, and $z$ is
the redshift of the GRB. We assumed isotropic emission.
Since the GRBs were detected by different instruments, their fluences (Col. 3)
refer to different energy ranges.  However, these ranges do not differ much,
and so we did not reduce the $\gamma$-ray fluxes to a common rest-frame energy
range. The properties of the GRB-SNe are summarized in Table 6.

This conundrum may be interpreted in two ways. One is that we may be seeing
an intrinsically similar phenomenon under different viewing angles.
GRB030329/SN2003dh may be viewed almost pole-on, while the
orientation of GRB980425/SN1998bw may be sufficiently off-axis
\citep[$\sim 15-30^{\circ}$,][]{mae02,mae06}
that the detected GRB is significantly weaker.
GRB031203/SN2003lw may lie somewhere in between \citep{err05}.
If this is indeed the case, then the $\gamma$-ray properties are a very
strong function of angle (\eg $E_{\gamma} \propto \theta^{-4})$.  In this
scenario, where line-of-sight inclination is a natural free parameter, the
optical properties of the SN are not much influenced by this relatively
small spread in viewing angles, since the asymmetry in the SN is much less
pronounced than it is in the GRB \citep{mae02}.
Larger spectral differences require larger inclination angles and are best
seen in the nebular phase \citep{maz05,mae06}

The other possibility is that there is a dispersion in the properties of the
relativistic ejecta for SNe with otherwise very similar characteristics. This
may well be, especially if we consider that the relativistic energies at play
in the GRB phenomenon ($\sim 10^{47}$ -- $10^{51}$\,erg) are small compared to
the kinetic energy involved in the SN event (from $\sim 10^{51}$\,erg for a
normal SN to $\sim 10^{52}$\,erg for a hypernova) and to the presumably even
larger neutrino energies ($\sim 10^{53}$ -- $10^{54}$\,erg), if there is a
bounce \citep{deng05}.

However, the inferred spread of properties of the SNe, albeit only of the order
of $\sim 30$\%, is seen not only in \KE\ and \Mej, but also in less
orientation-dependent quantities such as the mass of \Nifs, and therefore it is
almost certainly real. Most likely both intrinsic differences and orientation
effects are present.  So, while the observed $\gamma$-ray energy of GRB980425
was so low that it may require an intrinsically weak GRB with normal spectral
properties, the true energy may be higher than estimated since the GRB was
viewed off-axis as suggested by SN nebular line profile studies.

The case of GRB031203 is similarly uncertain: based on its X-ray to
$\gamma$-ray flux ratio, it was reported to be an XRF \citep{wat04,wat06a}.
However, \citet{saz04} suggested that it is a classical GRB based on the shape
of the $\gamma$-ray spectrum. Based on radio calorimetry and the absence of the
signature of an off-axis afterglow, \citet{saz04} and \citet{sod04} suggest
that it is an intrinsically underluminous event.  However, \citet{err05} model
the event as a standard-energy, off-axis GRB. Since nebular-phase spectra of
SN2003lw are not available, it is unfortunately impossible to estimate the
geometry of the explosion and the viewing angle.

Although it is now clear that long-duration GRBs and energetic, broad-lined
SNe~Ic (hypernovae) are related \citep{gal98,hjo03,mal04,pods04}, the physical
link between the relativistic event and these SNe remains uncertain. Further
uncontroversial occurrences of GRB-SN association will hopefully clarify the
issue.

Acknowledgements: This work was supported in part by the Grant-in-Aid for
Scientific Research (16540229, 17030005, and 17033002 for K.N.) and the 21st
Century COE Program (QUEST) of the JSPS and MEXT of Japan. We thank the referee
for a constructive report.




\clearpage

\begin{deluxetable}{lllccc}
\tabletypesize{\scriptsize}
\tablecaption{Late-time photometry.\label{tb:phot}}
\tablewidth{0pt}
\tablehead{
\colhead{Mean date (UT)} &
\colhead{Exposure time} &
\colhead{Seeing} &
\colhead{Filter} &
\colhead{Instrument} &
\colhead{Magnitude} \\
}
\startdata
2004 May 22.02 & 2$\times$2 min & 0.6\arcsec & $R$ &FORS\,1 & 20.44$\pm$0.02 \\
2004 May 22.02 & 2$\times$2 min & 0.6\arcsec & $I$ &FORS\,1 & 19.40$\pm$0.04 \\
\enddata
\end{deluxetable}


\begin{deluxetable}{lllcc}
\tabletypesize{\scriptsize}
\tablecaption{Spectroscopy of SN 2003lw.\label{tb:spec}}
\tablewidth{0pt}
\tablehead{
\colhead{Mean date (UT)} &
\colhead{Exposure time} &
\colhead{Seeing} &
\colhead{Grism} &
\colhead{Instrument}  \\
}
\startdata
2003 Dec 20.30 &2$\times$15 min &0.3\arcsec &150I  &FORS\,2    \\
2003 Dec 20.35 &2$\times$45 min &0.3\arcsec &300V  &FORS\,2    \\
2003 Dec 30.30 &2$\times$45 min &0.5\arcsec &300V  &FORS\,1    \\
2004 Mar 02.12 &4$\times$30 min &0.6\arcsec &300V  &FORS\,1    \\
2004 May 22.00 &3$\times$15 min &0.6\arcsec &150I  &FORS\,1    \\
\enddata
\end{deluxetable}


\begin{deluxetable}{ccccccc}
\tabletypesize{\scriptsize}
\tablecaption{Models for the December 20 spectrum}
\tablewidth{0pt}
\tablehead{
\colhead{Model} &
\colhead{$E_H(B-V)$} &
\colhead{$\log L$} &
\colhead{$v_{ph}$} &
\colhead{$V$} &
\colhead{M$_V$} &
\colhead{M$(Bol)$}\\
\colhead{} &
\colhead{} &
\colhead{[\ergs]} &
\colhead{\kms} &
\colhead{} &
\colhead{} &
\colhead{} \\
}
\startdata
1 & 0.20 & 43.01 & 17250 & 21.88 & -19.43 & -18.82 \\
2 & 0.25 & 43.09 & 18500 & 21.90 & -19.58 & -19.03 \\
3 & 0.30 & 43.16 & 18150 & 21.88 & -19.75 & -19.20 \\
4 & 0.35 & 43.26 & 19000 & 21.83 & -19.96 & -19.45 \\
\enddata
\end{deluxetable}


\begin{deluxetable}{ccccccc}
\tabletypesize{\scriptsize}
\tablecaption{Models for the December 30 spectrum}
\tablewidth{0pt}
\tablehead{
\colhead{Model} &
\colhead{$E_H(B-V)$} &
\colhead{$\log L$} &
\colhead{$v_{ph}$} &
\colhead{$V$}&
\colhead{M$_V$} &
\colhead{M$(Bol)$}\\
\colhead{} &
\colhead{} &
\colhead{[\ergs]} &
\colhead{\kms} &
\colhead{} &
\colhead{} &
\colhead{} \\
}
\startdata
1 & 0.20 & 42.85 & 10000 & 22.27 & -19.04 & -18.42 \\
2 & 0.25 & 42.91 & 10500 & 22.30 & -19.17 & -18.57 \\
3 & 0.30 & 43.00 & 11500 & 22.25 & -19.37 & -18.80 \\
4 & 0.35 & 43.09 & 10625 & 22.24 & -19.55 & -19.02 \\
\enddata
\end{deluxetable}


\begin{deluxetable}{ccrcccc}
\tabletypesize{\scriptsize}
\tablecaption{GRB Energetics}
\tablewidth{0pt}
\tablehead{
\colhead{GRB} &
\colhead{T} &
\colhead{$S_\gamma$} &
\colhead{Range} &
\colhead{Instrument} &
\colhead{Reference} &
\colhead{$E_{iso}$}\\
\colhead{} &
\colhead{s} &
\colhead{e-6 erg/cm2} &
\colhead{keV} &
\colhead{} &
\colhead{} &
\colhead{erg}\\
}
\startdata
980425 & 31 &   2.8$\pm$0.5 &  40-700 & SAX-GRBM     &  Pian et al. 2000     & $4.3 \times 10^{47}$ \\
       & 23 &   4.4$\pm$0.4 & 24-1820 & CGRO-BATSE   &  Galama et al. 1998   & \\
030329 & 63 &  55.3$\pm$0.3 &   2-30  & HETE-FREGATE &  Sakamoto et al. 2004 & \\
       & 63 & 107.6$\pm$1.4 &  30-400 & HETE-FREGATE &  Sakamoto et al. 2004 & $6.9 \times 10^{51}$ \\
       & 63 & 163.0$\pm$1.4 &   2-400 & HETE-FREGATE &  Sakamoto et al. 2004 & \\
031203 & 40 &   2.0$\pm$0.4 &  20-200 & INTEGRAL-IBIS & Sazonov et al. 2004  & $4.9 \times 10^{49}$ \\
\enddata
\end{deluxetable}



\begin{deluxetable}{ccccccc}
\tabletypesize{\scriptsize}
\tablecaption{Properties of GRB-SNe}
\tablewidth{0pt}
\tablehead{
\colhead{GRB/SN}        &
\colhead{\KE}   &
\colhead{M(\Nifs)}      &
\colhead{\Mej}          &
\colhead{M$_{\rm ZAMS}$} &
\colhead{Reference} \\
\colhead{}          &
\colhead{$10^{51}$ erg} &
\colhead{$\Msun$}   &
\colhead{$\Msun$}   &
\colhead{$\Msun$}   &
\colhead{}      \\
}
\startdata
GRB\,980425/SN\,1998bw & $50\pm 5$ & 0.38-0.48 & $10\pm1$ & 35-45 & Iwamoto \etal 1998, Nakamura \etal 2001, Maeda \etal 2003\\
GRB\,030329/SN\,2003dh & $40\pm10$ & 0.25-0.45 & $ 8\pm2$ & 25-40 & Mazzali \etal 2003, Deng \etal 2005\\
GRB\,031203/SN\,2003lw & $60\pm10$ & 0.45-0.65 & $13\pm2$ & 40-50 & this paper \\
\enddata
\end{deluxetable}


\onecolumn


\clearpage
\begin{figure}
\epsscale{0.80}
\plotone{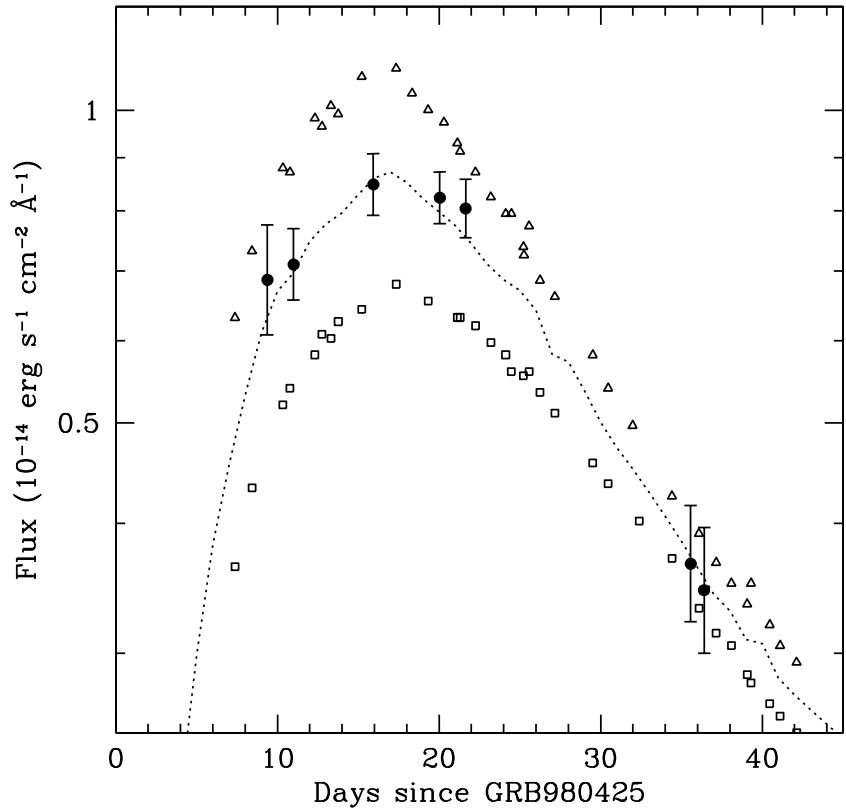}
\figcaption[f1.eps]{The $R$-band LC of SN~2003lw, galaxy-subtracted and
 time-stretched by a factor 0.9 (filled dots), compared to the $V$ (triangles)
 and $R$ (squares) LCs of SN~1998bw.
 Because of the redshift, the $R$-band LC of SN~2003lw is equivalent to
 rest-frame 6000\AA. The equivalent 6000\AA\ LC of SN~1998bw, obtained
 interpolating the $V$ and $R$ LCs, is shown as a dotted line.
 The LC of SN~2003lw has been shifted in flux by an arbitrary amount (a factor
 $900$) to match the corresponding 6000\AA\ LC of SN~1998bw.
 The shape of the "time compressed" light curve of SN~2003lw matches well that
 of SN~1998bw.
 \label{fig1}}
\end{figure}


\clearpage
\begin{figure}
\epsscale{1.0}
\plotone{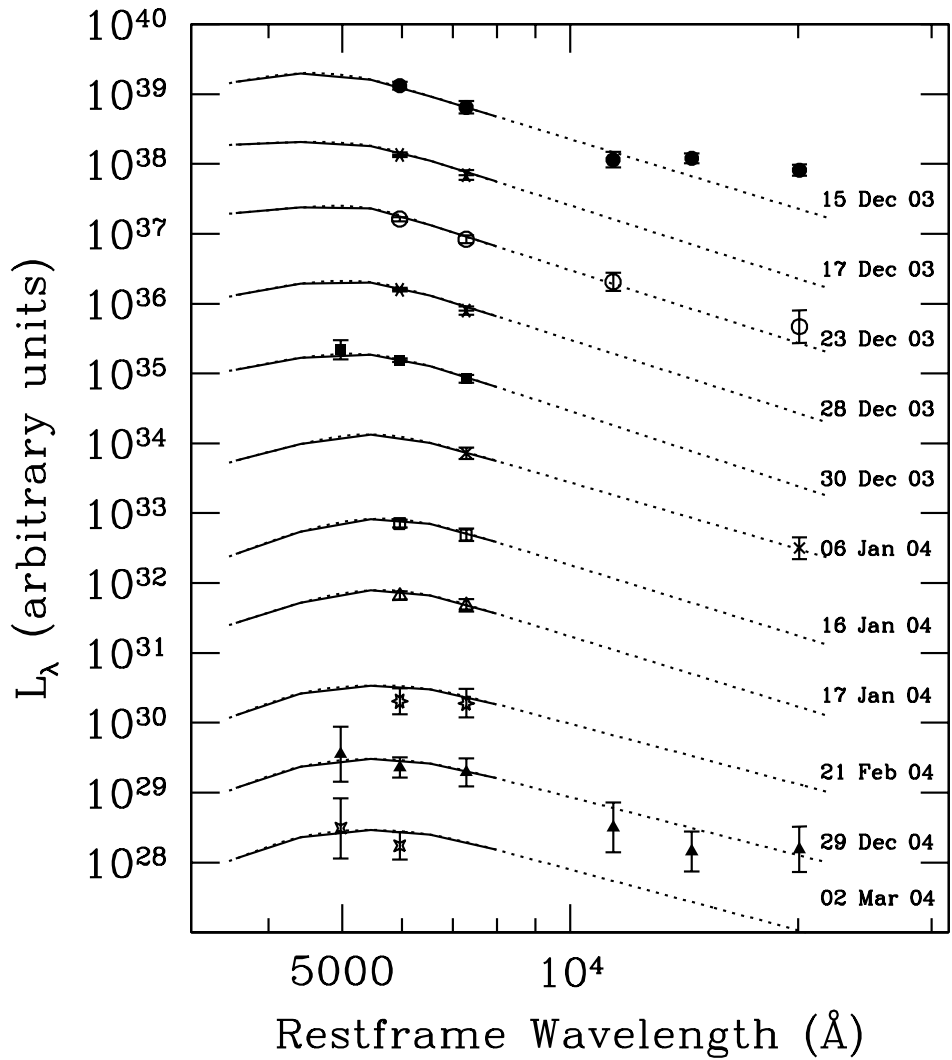}
\figcaption[f2.eps]{Broad-band spectra (SEDs) of SN2003lw corrected for
 Galactic and host reddening. These were computed
 using $H_0 = 72$\,\kms\,Mpc$^{-1}$ \citep{spe03} and a standard cosmology
 ($\Omega_{\Lambda} = 0.72$, $\Omega_{M} = 0.28$), which yields a distance to
 SN~2003lw of 490\,Mpc (\ie a distance modulus $\mu = 38.44$).  Only for the
 fluxes corresponding to the first epoch (top
 spectrum) the y-axis units are physical (erg s$^{-1}$ \AA$^{-1}$).
 The subsequent spectra, each represented by a different symbol, have been
 spaced apart by factors of 10, for clarity.  The solid curves show the UBVRI
 SEDs of SN~1998bw, dereddened for $E(B-V) = 0.016$ \citep{nak01},
 reconstructed at the phases
 corresponding to the epochs of the SN~2003lw spectra and multiplied by a
 constant factor of 1.3 to match the SN2003lw spectra. The dashed curves are
 the IR extrapolations of the SN~1998bw optical spectra. \label{fig2}}
\end{figure}


\clearpage
\begin{figure}
\plotone{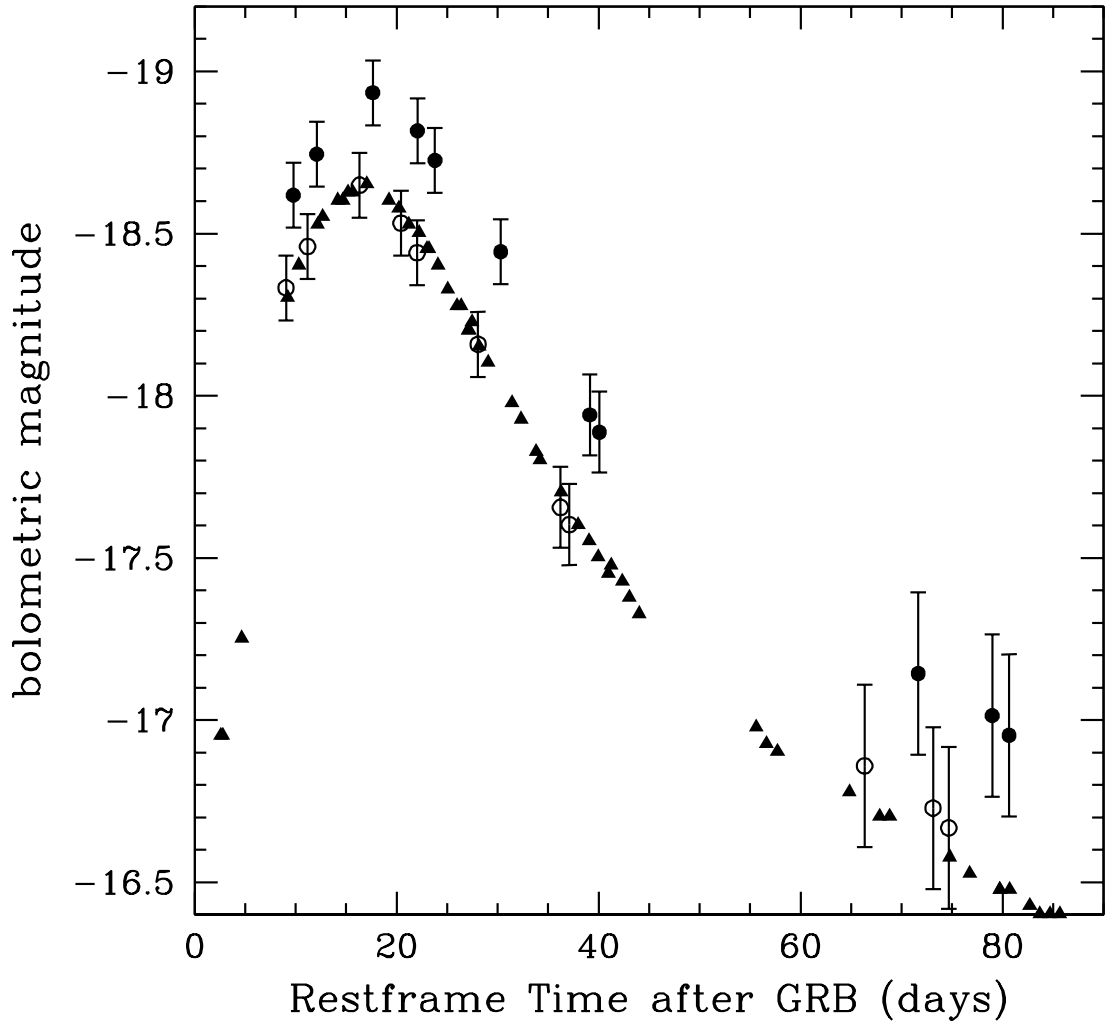}
\figcaption[f3.eps]{The {\em ``uvoir''} bolometric light curves of SNe~2003lw
 (filled dots) and 1998bw (triangles). The LC of SN~1998bw was computed for a
 distance modulus $\mu=32.76$ and $E(B-V) = 0.016$, that of SN~2003lw for
 $\mu=38.44$, $E_G(B-V) = 0.8$, and $E_H(B-V) = 0.25$. The open circles show
 the LC of SN~2003lw stretched by a factor of 0.9 and shifted in flux to match
 the LC of SN~1998bw. Error bars do not include a sistematic uncertainty of $\pm
 0.2$\,mag (Section 3) \label{fig3}}
\end{figure}


\clearpage
\begin{figure}
\plotone{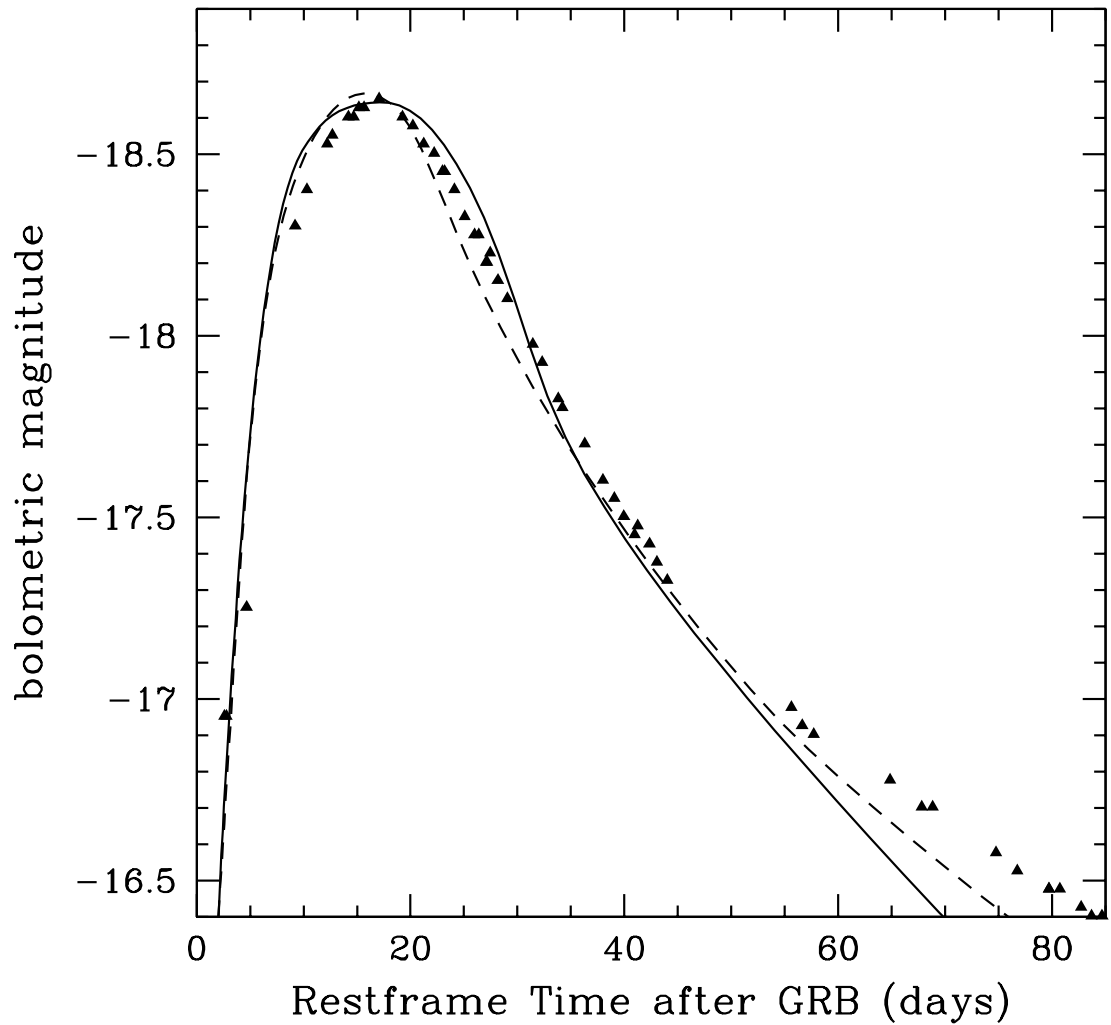}
\figcaption[f4.eps]{Models for the {\em ``uvoir''} bolometric light curve of
 SN~1998bw (triangles): the continuous line is model CO138E50, the dashed line
 is the 2-component model. \label{fig4}}
\end{figure}


\clearpage
\begin{figure}
\plotone{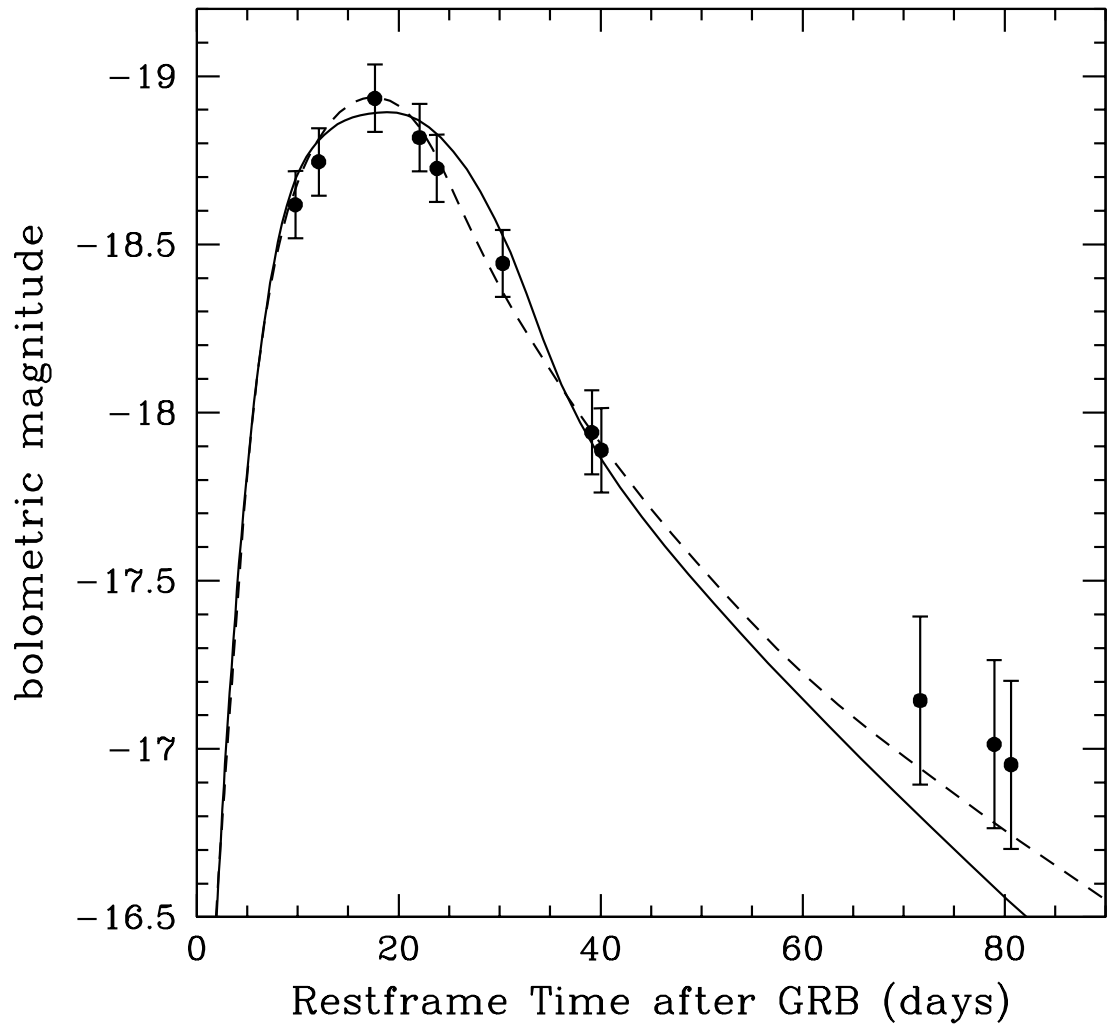}
\figcaption[f5.eps]{Models for the {\em ``uvoir''} bolometric light curve of
 SN~2003lw (dots): the continuous line is the rescaled model CO138E50, the
 dashed line is the 2-component model. \label{fig5}}
\end{figure}


\clearpage
\begin{figure}
\epsscale{1.0}
\plotone{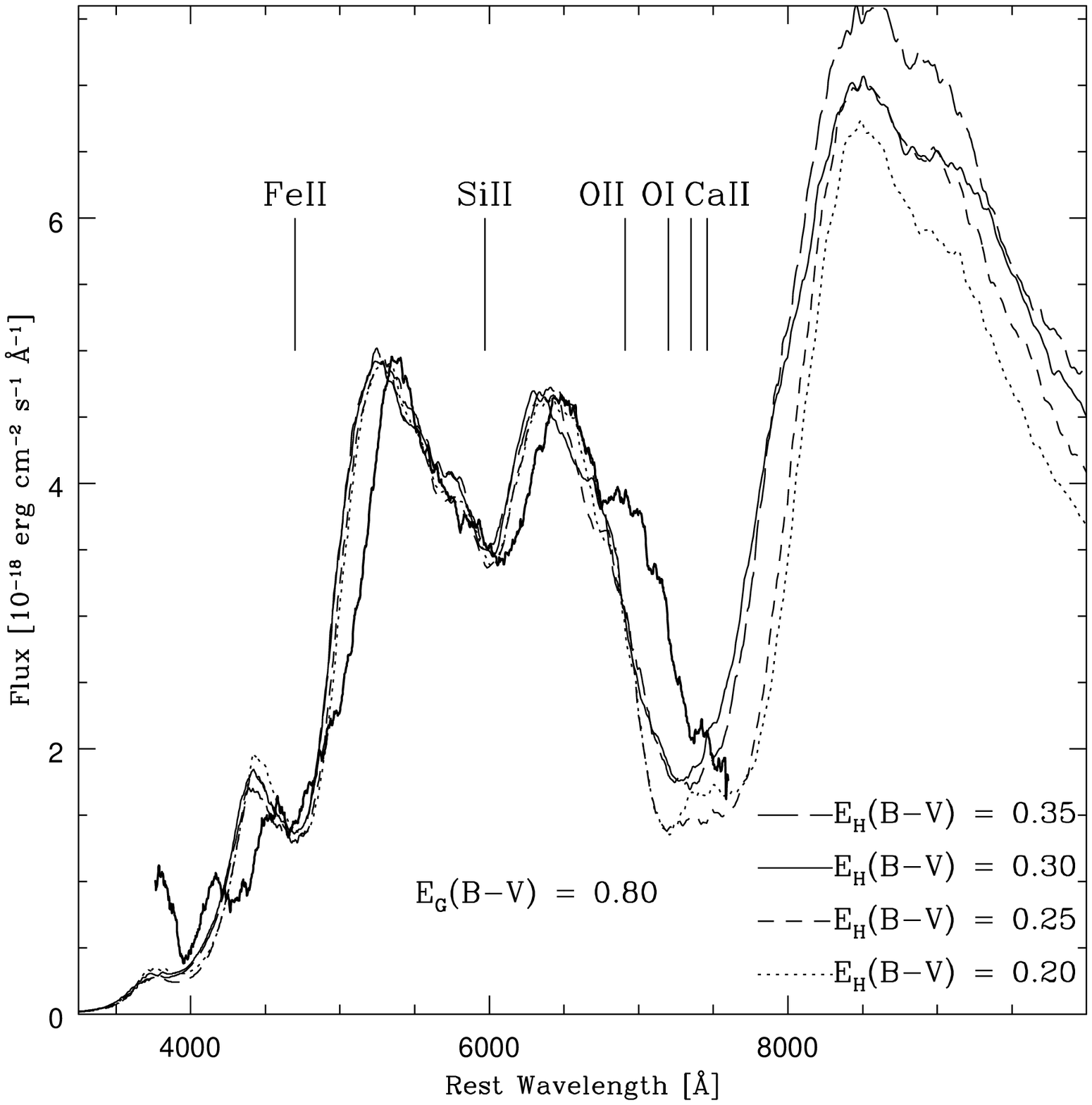}
\figcaption[sn03lw20Decredd080xBW.eps]{The VLT spectrum of SN~2003lw obtained on
 December 20, 2003 (thick line), compared to four synthetic spectra, computed
 with $E_G(B-V) = 0.8$ and $E_H(B-V) = 0.2$ (dotted line), 0.25 (dashed line),
 0.3 (thin continuous line), and 0.35 (long-dashed line), respectively.
 \label{fig6}}
\end{figure}


\clearpage
\begin{figure}
\plotone{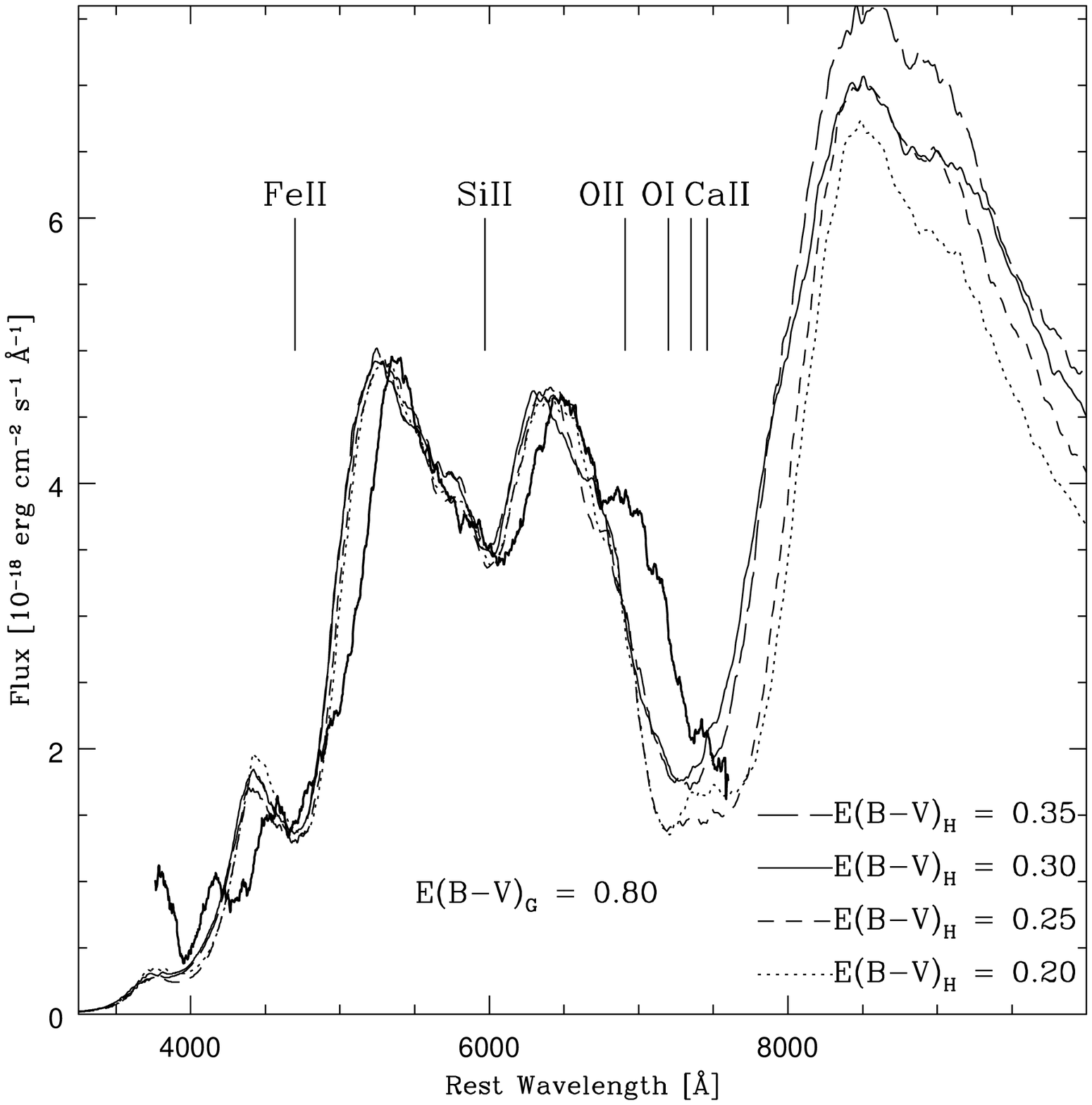}
\figcaption[sn03lw30Decredd080xBW.eps]{The VLT spectrum of SN~2003lw obtained on
 December 30, 2003 (thick line), compared to four synthetic spectra, computed
 with $E_G(B-V) = 0.8$ and $E_H(B-V) = 0.2$ (dotted line), 0.25 (short-dashed
 line), 0.3 (thin continuous line), and 0.35 (long-dashed line), respectively.
 \label{fig7} }
\end{figure}


\clearpage
\begin{figure}
\plotone{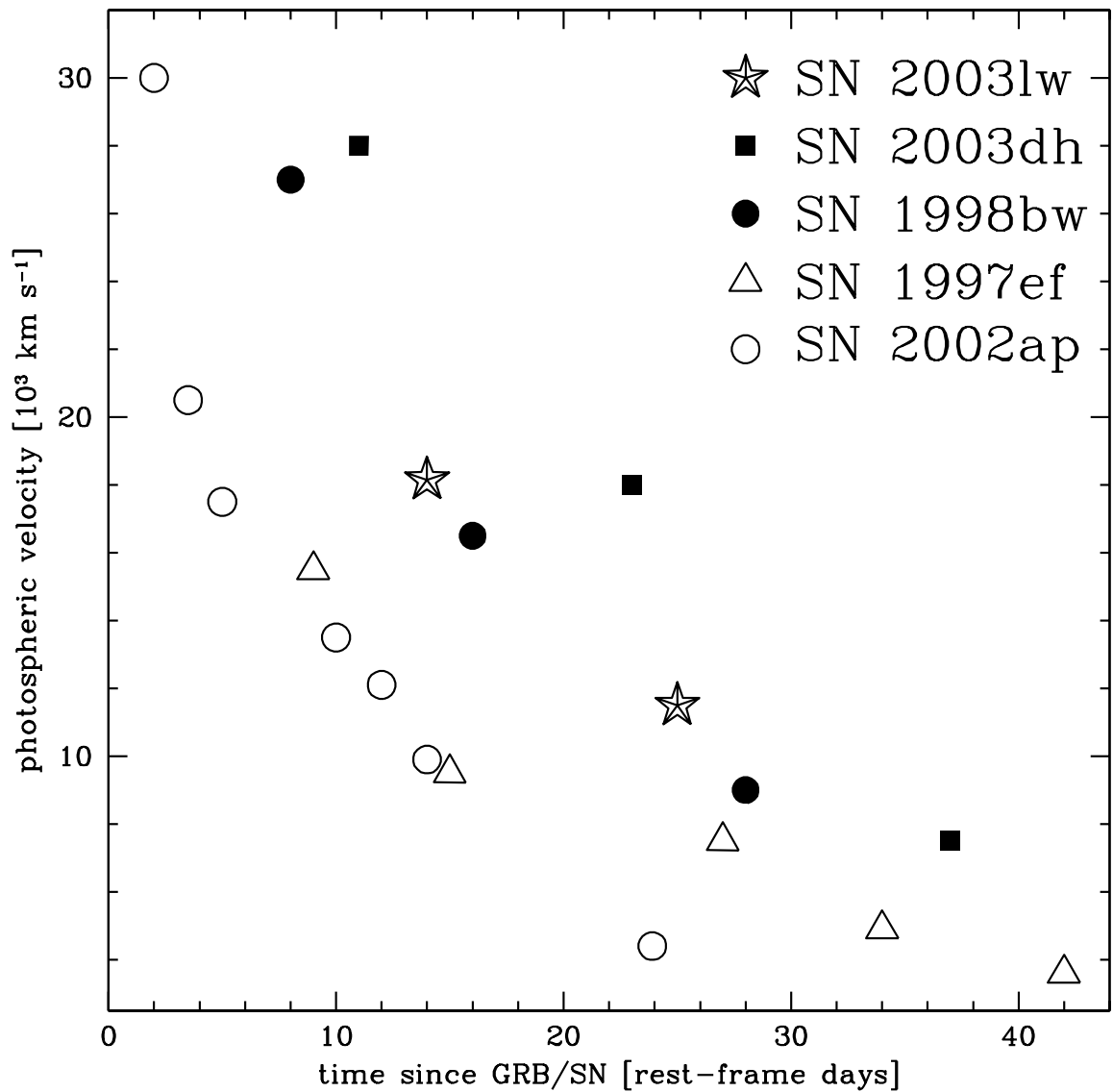}
\figcaption[HNvel.eps]{Evolution of the photospheric velocity in SN~2003lw
 (stars) and in a number of broad-lined SNe~Ic. The references for the data are:
 SN~1997ef: \citet{iwa00}; SN~1998bw: \citet{iwa98}; SN~2002ap: \citet{maz02};
 SN~2003dh: \citet{maz04}.
 \label{fig8}}
\end{figure}


\end{document}